\documentclass[twocolumn,aps,prl,showpacs,superscriptaddress]{revtex4}

\pdfoutput=1
\usepackage{graphicx}
\usepackage{amsmath,amssymb}
\usepackage{color}
\usepackage{multirow}
\usepackage[latin1]{inputenc}
\usepackage[T1]{fontenc}
\usepackage[english]{babel}
\usepackage{amsmath,amssymb,amsthm}
\usepackage{hyperref}
\usepackage{booktabs}
\usepackage{bbm}
\usepackage{url}
\usepackage{nicefrac}
\usepackage{multirow} 
\usepackage{graphicx} 
\usepackage[normalem]{ulem}

\newcommand{\brc}[1]{\left\{ #1 \right\}}

\newcommand{\ket}[1]{\left| #1 \right\rangle}
\newcommand{\braket}[2]{\left\langle {#1{\left| \vphantom{#1 #2} \right.} #2} \right\rangle}

\usepackage{array}
\newcolumntype{C}{>{\(}c<{\)}}
\newcolumntype{L}{>{\(}l<{\)}}
\renewcommand{\epsilon}{\varepsilon}
\renewcommand{\phi}{\varphi}

\begin{document}

\title{Test of mutually unbiased bases for six-dimensional photonic quantum systems}
%
\author{Vincenzo D'Ambrosio}
\affiliation{Dipartimento di Fisica, Sapienza Universit\`{a} di Roma, Roma 00185, Italy}
\author{Filippo Cardano}
\affiliation{Dipartimento di Fisica, Universit\`{a} di Napoli ``Federico II'', Compl.\ Univ.\ di Monte S. Angelo, 80126 Napoli, Italy}
\author{Ebrahim Karimi}
\email{ekarimi@uottawa.ca}
\altaffiliation[Current address: ]{Department of Physics, University of Ottawa, 150 Louis Pasteur, Ottawa, Ontario, K1N 6N5 Canada}
\affiliation{Dipartimento di Fisica, Universit\`{a} di Napoli ``Federico II'', Compl.\ Univ.\ di Monte S. Angelo, 80126 Napoli, Italy}
\author{Eleonora Nagali}
\affiliation{Dipartimento di Fisica, Sapienza Universit\`{a} di Roma, Roma 00185, Italy}
\author{Enrico Santamato}
\affiliation{Dipartimento di Fisica, Universit\`{a} di Napoli ``Federico II'', Compl.\ Univ.\ di Monte S. Angelo, 80126 Napoli, Italy}
\affiliation{Consorzio Nazionale Interuniversitario per le Scienze Fisiche della Materia, Napoli}
\author{Lorenzo Marrucci}
\email{marrucci@na.infn.it}
\affiliation{Dipartimento di Fisica, Universit\`{a} di Napoli ``Federico II'', Compl.\ Univ.\ di Monte S. Angelo, 80126 Napoli, Italy}
\affiliation{CNR-SPIN, Compl.\ Univ.\ di Monte S. Angelo, 80126 Napoli, Italy}
\author{Fabio Sciarrino}
\email{fabio.sciarrino@uniroma1.it}
\affiliation{Dipartimento di Fisica, Sapienza Universit\`{a} di Roma, Roma 00185, Italy}

\begin{abstract}
In quantum information, complementarity of quantum mechanical observables plays a key role. If a system resides in an eigenstate of an observable, the probability distribution for the values of a complementary observable is flat. The eigenstates of these two observables form a pair of mutually unbiased bases (MUBs). More generally, a set of MUBs consists of bases that are all pairwise unbiased. Except for specific dimensions of the Hilbert space, the maximal sets of MUBs are unknown in general. Even for a dimension as low as six,  the identification of a maximal set of MUBs remains an open problem, although there is strong numerical evidence that no more than three simultaneous MUBs do exist. Here, by exploiting a newly developed holographic technique, we implement and test different sets of three MUBs for a single photon six-dimensional quantum state (a ``qusix''),  encoded either in a hybrid polarization-orbital angular momentum or a pure orbital angular momentum Hilbert space. A close agreement is observed between theory and experiments. Our results can find applications in state tomography, quantitative wave-particle duality, quantum key distribution and tests on complementarity and logical indeterminacy.
\end{abstract}

\maketitle

\section*{INTRODUCTION}
The ``complementarity'' of different observables of a same physical system is one of the basic features of the quantum world \cite{Bohr}. Besides its fundamental relationship with the concept of wave-particle duality, complementarity today plays a key role in areas ranging from quantum state reconstruction \cite{Wootters:1989} to quantum information applications such as quantum key distribution \cite{Gisi02}. Mathematically, complementary observables are described by noncommuting Hermitian operators whose sets of eigenstates form different bases in the Hilbert space that are said to be ``mutually unbiased'' (MUBs). This expression refers to the fact that the overlap (or inner product) of any pair of states belonging to different bases is the same \cite{Durt:2010}.

In quantum cryptography, complementary observables and the associated MUBs are the core of all protocols proposed for secure quantum key distribution, starting from the famous BB84 protocol \cite{Benn84} and its extension to three qubit bases \cite{Pasq99}. The ``no cloning theorem'' implies that Alice and Bob can always recognize a possible eavesdropper attack by detecting the associated disturbance introduced in the system. In particular, the adoption of MUBs for encoding the information is known to maximize this disturbance allowing one to recognize the attack most effectively \cite{Gisi02}.

In high-dimensional systems, complementary observables and the corresponding MUBs have been exploited to enhance the security in quantum cryptograpy \cite{Cabe11}, perform fundamental tests of quantum mechanics, such as quantum contextuality \cite{Koch67,Naga12,Damb13}, explore logical indeterminacy \cite{Bruk09}, and several other tasks in quantum information. For instance, new quantum key distribution protocols were conceived in which a larger error rate can be tolerated while preserving security \cite{Cerf02,Caru05}. Moreover a different protocol extending Ekert91 \cite{Eker91} by using entangled qutrits has been experimentally realized \cite{Grob06}. In quantum state tomography, MUBs play a crucial role because they correspond to the optimal choice of the measurements to be performed in order to obtain a full reconstruction of the density matrix.

Given a Hilbert space of dimension $d$, an important problem is to find the maximum number of MUBs that can be defined simultaneously. Although for spaces of prime-power dimensions there exist several methods to find a maximal set of $d+1$ MUBs \cite{Plan06}, 
this problem remains in general hitherto unsolved \cite{Band02,Beng06}. Dimension six, in particular, has been widely investigated in the last few years \cite{Beng07,Brierley:2008,Brierley:2009a,Rayn11} because it is the lowest one for which the problem is still open. Nevertheless, several theorems imposing restrictions on the properties of the maximal set of MUBs have been proved \cite{McNu12,Weig12} and strong numerical evidence suggests that no more than three mutually unbiased bases actually exist in dimension six \cite{Brierley:2008,Rayn11,Butt07}.

Different experimental approaches have been recently adopted to implement complete sets of MUBs for state reconstruction in photonic systems. For example, the polarization of a photon pair was used to define MUBs in dimension four \cite{Stei10}, the orbital angular momentum (OAM) of single photons has been used to address Hilbert spaces with $d=2,3,4,5$ \cite{Giov12}, and multiple propagation modes were combined to reach dimensions $d=7,8$ \cite{Lima11}. Hybrid methods combining polarization and OAM were also used to define and manipulate photonic ququarts ($d=4$) \cite{Naga10,Naga10b}. However, since in $d=6$ no complete set of MUBs is known, this case has not been investigated hitherto for state tomography and, to our knowledge, even the minimal set of three MUBs has never been demonstrated in an experimental framework. In the present work, we shall focus our attention on this important six-dimensional (6D) case.

In this paper we demonstrate the generation of MUBs in 6D by exploiting two different approaches. A 6D Hilbert space can be always decomposed in the tensor product of a two-dimensional (2D) and a three-dimensional (3D) space. Hence, a possible route to implementing 6D quantum states (qusix) is by combining the 2D and 3D spaces related to two different degrees of freedom of the photon. Following this route, in our first experiment we prepare and analyze all states of three MUBs in a 6D Hilbert space obtained by combining the 2D space of polarization and a given 3D subspace of the OAM. The OAM degree of freedom is related to the photon's transverse-mode spatial structure, and in principle it allows one to define a single-photon Hilbert space of arbitrarily large dimensionality \cite{Alle92,Moli08,Fran08,Marr11}. In our second experiment, we exploit this feature to prepare and test three MUBs in a 6D space defined only in the OAM of the photon. The polarization qubit is easily manipulated with standard optical elements, while arbitrary OAM states can be obtained by computer-generated holography using a spatial light modulator (SLM). In the present work we developed a novel method to determine the hologram kinoform in order to obtain a sufficiently high fidelity in the state generation. Indeed, MUBs are very sensitive to the generation fidelity and relatively small imperfections are immediately visible in the MUBs cross overlaps. A high fidelity is also crucial to exploit MUBs for quantum cryptography.

\section{MUTUALLY UNBIASED BASES}
Let \(A\) and \(B\) be two operators in a \(d\)-dimensional Hilbert space, with orthonormal eigenbases \(\brc{\ket{a_i}}\) and \(\brc{\ket{b_i}}\) respectively. Eigenstates of these observables are said to be \emph{mutually unbiased} \cite{Wootters:1989, Durt:2010} if
\begin{equation}
|\braket{a_i}{b_j}|^2=\frac{1}{d},\quad \forall \,i,j\in\{1,. . .,d\}.
\end{equation}
Such operators are also called mutually complementary, or maximally noncommutative, since given any eigenstate of one, the outcome  resulting from a measurement of the other is completely undetermined. In a $d$-dimensional Hilbert space a pair of MUBs can always be found. Indeed, let $\brc{\ket{a_i}}$ be the computational basis,
\begin{equation}
\brc{\ket{a_i}}=\brc{\ket{0},\ket{1},\mbox{...},\ket{d-1}}.
\end{equation}
A discrete Fourier transform can be then used to define the following dual basis, which is mutually unbiased to the previous one:
\begin{equation}
\ket{b_i}=\frac{1}{\sqrt d}\sum_{j=0}^{d-1}\omega_d^{\,ij}\ket{a_j}
\label{trasfbasis}
\end{equation}
where $\omega_d=\exp{\left(\text{i} 2\pi/d \right)}$, and the non-italic $\text{i}$ denotes the imaginary unit (not to be confused with the index $i$). The pair of operators associated to these bases, often named $\hat{Z}$ and $\hat{X}$ reminiscent of the Pauli operators, provides an algebraic complete set of observables that fully parametrizes the physical degree of freedom described by the Hilbert space: all other operators acting on this space are product of powers of $\hat{Z}$ and $\hat{X}$ \cite{Durt:2010}.

An open issue concerns the maximal number of MUBs that can be found in a $d$-dimensional space; in the specific case when $d$ is equal to a prime number or to a prime power, a maximal set of $d+1$ MUBs does exist \cite{Durt:2010}. This set is also ``complete'', in the sense that by projective measurements over its states  $(d-1)(d+1)=d^2-1$ independent real parameters can be obtained, which are exactly the number of parameters needed for full density matrix reconstruction \cite{Wootters:1989}. A complete set of MUBs can be found using several methods, i.e., the Galois Field, the Heisenberg-Weyl group, Hadamard matrices, etc.\ (for a review see \cite{Durt:2010,Plan06}). However, in the general case of composite dimensions that are not prime powers such as $d=6,10,12,...$, all these methods fail \cite{Arch03}. On the base of extensive numerical simulations, it has been conjectured that complete sets of MUBs do not exist in this case \cite{Butt07}, although such conjecture hitherto has not been rigorously proven. A minimum number of MUBs that is known to exist in such cases is given by $p^k+1$, where $p^k$ is the lowest factor in the prime decomposition of the number $d$ \cite{Klappenecker:2004}. For instance, in the $d=6$ case, three MUBs can be easily constructed, but no evidence for the existence of a fourth basis that is unbiased with the first three has ever been found.

The Hilbert space $\mathcal H^6$ of a 6D system can be always factorized in the direct product of a 2D and a 3D space, i.e.,
\begin{equation}
\mathcal H^6=\mathcal H^2 \otimes \mathcal H^3.
\end{equation}
Both these Hilbert subspaces possess complete sets of MUBs, $\brc{\ket{m_i^\alpha}}$ and $\brc{\ket{n_j^\beta}}$, containing respectively three and four bases. Although the states of these bases can be combined to form twelve different separable bases for the space $\mathcal H^6$, which can be used for a complete tomography of the qusix state, only sets with three MUBs can be constructed. A possible choice is given by the following three bases:
\begin{eqnarray} \label{combmubs}
I=\{ \ket{m^1_i}\otimes\ket{n^1_j} \} \nonumber\\
II=\{ \ket{m^2_i}\otimes\ket{n^2_j} \} \\
III=\{ \ket{m^3_i}\otimes\ket{n^3_j} \} \nonumber
\end{eqnarray}
where $i\in\brc{1,2}$ and $j\in\brc{1,2,3}$; an explicit matricial expression of states that we consider is reported in the Methods section. It can be immediately seen that any other basis obtained introducing the fourth one $\brc{\ket{n^4_j}}$ of the space $\mathcal{H}_3$ would not be mutually unbiased with the others, since it is missing a different basis in $\mathcal{H}_2$. This set of 18 product states cannot be extended by any other vector in $\mathcal{H}_6$, even if entangled states are considered \cite{McNu12}; moreover if a complete MUB set in $d=6$ existed, then only one among the seven bases therein could be composed of product states, while all others must be entangled \cite{Weig12}.

\subsection*{OAM ENCODING AND THE HOLOGRAPHIC TECHNIQUE}
Let $\ket{m}$ denote the eigenstates of the photon OAM with eigenvalue $m\hbar$, where $m$ is any integer. These states define the entire infinite-dimensional Hilbert space of OAM. We note that a full specification of the transverse optical modes would actually require assigning also a radial number, as in the case of Laguerre-Gauss modes. However, here and in the following we will omit this radial number and the specification of a given radial profile will be understood for each value of $|m|$.

\begin{figure}[htbp]
\includegraphics[width=8.5cm]{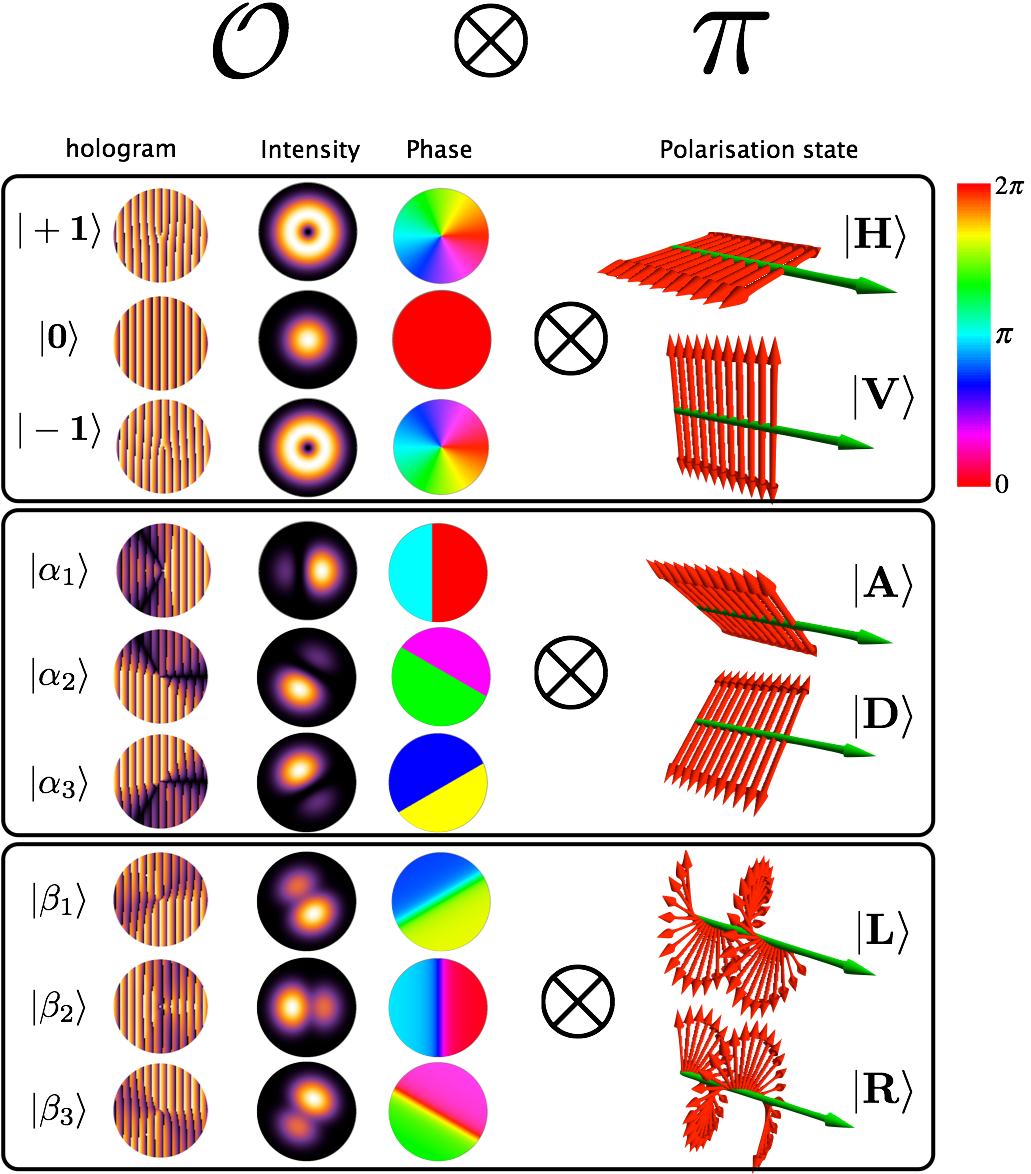}
\caption{\textbf{MUBs for hybrid photonic qusix encoding:} Representation of quantum states with dimension $d=6$ obtained from the direct product of a three-dimensional subspace $\mathcal{O}$ of OAM and the two-dimensional space $\pi$ of polarization. The three main boxes correspond to the three MUBs. On the left side, the intensity and phase distributions of each OAM spatial mode and the corresponding generating kinoform are shown. On the right side the polarization states are illustrated graphically by showing the optical electric field orientation in space at a given time.}
\label{state}
\end{figure}
We now define a 3D subspace $\mathcal{O}$ in OAM as that spanned by the three eigenvectors $\{\ket{+1},\ket{0},\ket{-1}\}$. These states can be taken to define the logical basis $\mathcal{O}_1$ of a photonic qutrit in the Hilbert space $\mathcal{O}$. A second basis in $\mathcal{O}$ that is mutually unbiased with the logical one can be obtained as the Fourier-transform one $\mathcal{O}_2 = \{\ket{\alpha_1}, \ket{\alpha_2}, \ket{\alpha_3}\}$ defined as in Eq.\ (\ref{trasfbasis}). Explicitly, we have $\ket{\alpha_1}=(\ket{-1}+\ket{0}+\ket{1})/\sqrt{3}, \ket{\alpha_2}=(\ket{-1}+\omega\ket{0}+\omega^2\ket{+1})/\sqrt{3}, \ket{\alpha_3}=(\ket{-1}+\omega^2\ket{0}+\omega\ket{+1})/\sqrt{3}$, where $\omega=\omega_3=\exp{\left(\text{i} 2\pi/3 \right)}$ and we have used the identity $\omega^4=\omega$. The other two bases of a maximal set of MUBs in $\mathcal{O}$ are then defined as follows: $\mathcal{O}_3 = \{\ket{\beta_1}, \ket{\beta_2}, \ket{\beta_3}\}$ with $\ket{\beta_1}=(\ket{-1}+\omega\ket{0}+\omega\ket{1})/\sqrt{3}, \ket{\beta_2}=(\ket{-1}+\omega^2\ket{0}+\ket{+1})/\sqrt{3}, \ket{\beta_3}=(\ket{-1}+\ket{0}+\omega^2\ket{+1})/\sqrt{3}$ and $\mathcal{O}_4 = \{\ket{\gamma_1}, \ket{\gamma_2}, \ket{\gamma_3}\}$ with $\ket{\gamma_1}=(\ket{-1}+\omega^2\ket{0}+\omega^2\ket{1})/\sqrt{3}, \ket{\gamma_2}=(\ket{-1}+\omega\ket{0}+\ket{+1})/\sqrt{3}, \ket{\gamma_3}=(\ket{-1}+\ket{0}+\omega\ket{+1})/\sqrt{3}$.

As shown in Eq.\ (\ref{combmubs}), in order to construct three MUBs in the 6D hybrid space, we will only need the first three bases $\mathcal{O}_1, \mathcal{O}_2, \mathcal{O}_3$. The intensity and phase profiles of the nine OAM states belonging to these three MUBs are shown in Fig.\ \ref{state}.

Let us now discuss the experimental method we adopted for the generation (and detection) of these nine states, and of all other OAM superposition states in this work. Arbitrary optical field transverse modes can be obtained by diffraction of an input Gaussian TEM$_{00}$ mode on a SLM programmed for displaying a prescribed ``kinoform'', that is the pattern determining the phase retardation experienced by the input wave in diffraction. The main problem is that the SLM is a phase-only optical element, while to obtain arbitrary OAM modes we need to be able to tailor both the phase and the amplitude transverse profiles of the outgoing field. This can be obtained by modulating both the shape and contrast of the kinoform fringes. To determine the kinoform, we initially tried some of the most commonly used methods \cite{arrizon07}, but found that they often give rise to a non-negligible ``cross-talk'', i.e., nonzero overlaps between different states of the same basis, and to significant unbalances in the overlaps of each state of a given basis with the states of other bases. In other words, the generation fidelity of these methods was not good enough for our purposes. For this reason, we developed an holographic method that is specifically optimized in the fidelity.

Let us first assume that the input field is a plane wave. Our goal is to obtain in the first order of diffraction a prescribed optical field ${\cal A} e^{i\Phi}$, $\cal{A}$ and $\Phi$ being the optical field normalized amplitude and phase, respectively. A straightforward calculation \cite{newpaper} shows that such optical field is obtained in the far field if the kinoform phase modulation has the following expression: 
\begin{equation}\label{eq:technique}
	\mathcal{M}=\text{Mod}\left[\left(\Phi-\pi {\cal I}+\frac{2\pi x}{\Lambda} \right),2\pi\right] \, {\cal I}
\end{equation}
where $\Lambda$ is the grating period that fixes the diffraction angle, $\mathcal{I}=\left(1+\text{sinc}^{-1}({\cal A})/\pi\right)$, in which $\text{sinc}^{-1}$ stands for inverse of $\text{sinc}(x)=\sin(x)/x$ function in the domain $[-\pi,0]$, and $\text{Mod}$ is the modulo function that gives the remainder after division of the first argument by the second. The inverse of $\text{sinc}$ function was evaluated numerically by the Newton method, with an accuracy of seven digits.

By this method we calculated the kinoforms needed to generate the nine OAM states of the first three MUBs of the OAM qutrit. The resulting hologram patterns are shown in Fig.\ \ref{state}. It can be seen that these kinoforms include only an azimuthal dependence, since the OAM state definition ignores the radial coordinate. This implies that the same kinoforms can also be used with a Gaussian input beam instead of a plane wave and only the radial profile of the diffracted wave will be affected, while the OAM state will remain the same. Moreover, we do not need to finely adjust the input beam waist of the Gaussian beam.

We note that the holograms defined by Eq.\ (\ref{eq:technique}) generate ideally exact modes in the far field, so that the expected overlap between states belonging to the same basis vanishes identically and that between states belonging to different MUBs is $1/3$ in the qutrit space (and hence it will be $1/6$ in the qusix space, after combining with polarization). As mentioned, this is not the case for other commonly used holographic methods. For example, numerical simulations based on the method reported in Ref. \cite{Leach05} yielded mean state fidelities of 100\%, 88.5\% and 84.5\% for the three OAM qutrit bases. This corresponds to 6\% and 7.7\% of mean cross-talk between states belonging to the last two bases. Moreover, the overlap between states of different MUBs is found to vary between 21\% and 48\%, depending on the state pair. These fidelity problems are absent in our method. More details about the performances of the holographic method used in this work will be reported elsewhere  \cite{newpaper}.

\subsection*{HYBRID QUSIX ENCODING AND CHARACTERIZATION}
Our first experimental implementation of qusix photonic states has been achieved by combining the 2D space $\pi$ of polarization and the 3D subspace $\mathcal{O}$ of OAM in single photons. The logical basis $I$ of quantum states in dimension six has been hence implemented as follows:
\begin{equation}\label{logicbase}
I=\{\ket{H,-1},\ket{H,0},\ket{H,+1},\ket{V,-1},\ket{V,0},\ket{V,+1}\}
\end{equation}
where $\ket{H},\ket{V}$ denote horizontal and vertical linear polarizations, as shown in Fig.\ \ref{state}, which have been combined with the three eigenstates of OAM forming the basis $\mathcal{O}_1$.

Following Eq.\ (\ref{combmubs}), a second basis $II$, unbiased with the first, is obtained by combining the diagonal/antidiagonal polarization states $\{\ket{A}=\frac{1}{\sqrt{2}}(\ket{H}+\ket{V}),\ket{D}=\frac{1}{\sqrt{2}}(\ket{H}-\ket{V})\}$ with the OAM states of basis $\mathcal{O}_2$. The third basis of the set of MUBs was finally obtained by combining the circular polarization states $\{\ket{L}=\frac{1}{\sqrt{2}}(\ket{H}+i\ket{V}$),$\ket{R}=\frac{1}{\sqrt{2}}(\ket{H}-i\ket{V})\}$ with the OAM third basis $\mathcal{O}_3$.

%
In order to experimentally generate these hybrid qusix states we employed the setup shown in Fig.\ \ref{setup}. Single photons emitted at a 7 kHz rate via spontaneous parametric down conversion (SPDC) in a beta-barium-borate (BBO) crystal \cite{Dema05} are collected by a single-mode fiber (SM) to filter out all the spatial modes but the Gaussian mode TEM$_{00}$, i.e., OAM state $\ket{0}$ (OAM qutrit initialization). A set of waveplates (C) compensates the polarization after the transmission through the fiber. The photons are then sent through a polarizing beam splitter (PBS) (polarization qubit initialization) and, after adjusting the radial-mode size by a pair of lenses (MA), to a first reflecting spatial light modulator (SLM1) which generates the desired OAM qutrit state. The hologram kinoform displayed on the SLM1 for each OAM state to be generated, in the first-order diffraction, is shown in Fig.\ \ref{state}. After SLM1, a half wave-plate (HWP) and a quarter wave-plate (QWP)  are used to write the polarization qubit in the photon. Hence we are able to generate any hybrid qusix that is a product of a qutrit and a qubit.

The qusix-carrying photon is then sent to the detection stage. This stage is composed of a polarization analysis set (HWP, QWP and a PBS) and a second spatial light modulator (SLM2) for converting in diffraction the OAM state to be detected back into a Gaussian mode. The photon is finally coupled to a single mode fiber, to filter only this Gaussian mode, connected to a single-photon counter module (SPCM). To eliminate the Gouy phase-shift effects between different OAM eigenstates occurring in free propagation, an imaging system (not shown in the figure) has been included to image the screen of SLM1 onto the SLM2. All waveplates and SLMs were computer-controlled so as to allow for a fully automatic generation and measurement procedure. With this setup, it is possible to perform a projective measurement upon every possible separable state of polarization and OAM.

\begin{figure}[htbp]
\includegraphics[scale=.21]{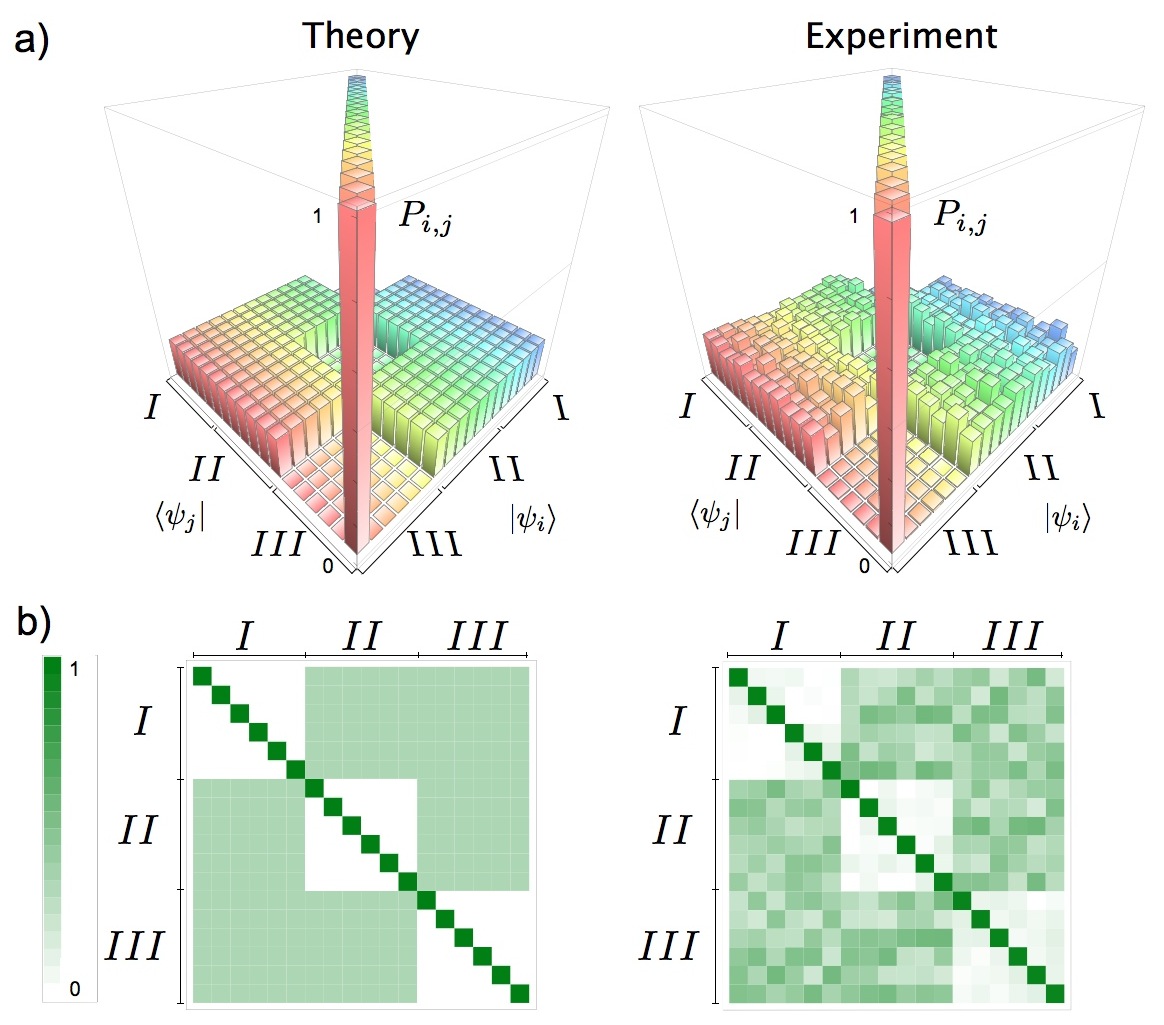}
\caption{\textbf{Experimental analysis of hybrid qusix photonic states:} Probability distribution resulting from all $18\times18$ projections of each state within the three MUBs over all the others, comparing theoretical and experimental values. According to theoretical predictions, we expect  that the $18\times18$ matrix can be divided into nine $6\times6$ blocks $A^m_n$, where the two indices $m,n\in\brc{\mbox{I,II,III}}$ label generation and detection bases, respectively. Blocks that correspond to projection of one basis over itself ($m=n$) should be diagonal, i.e., $(A^m_m)_{i,j}=\delta_{ij}$. Other blocks, whose values represent the overlap between states belonging to two different bases, should be flat, i.e., $(A^m_n)_{i,j}=1/6$, for $m\neq n$.} 
\label{hybrid1}
\end{figure}

\begin{figure*}
\includegraphics[scale=.25]{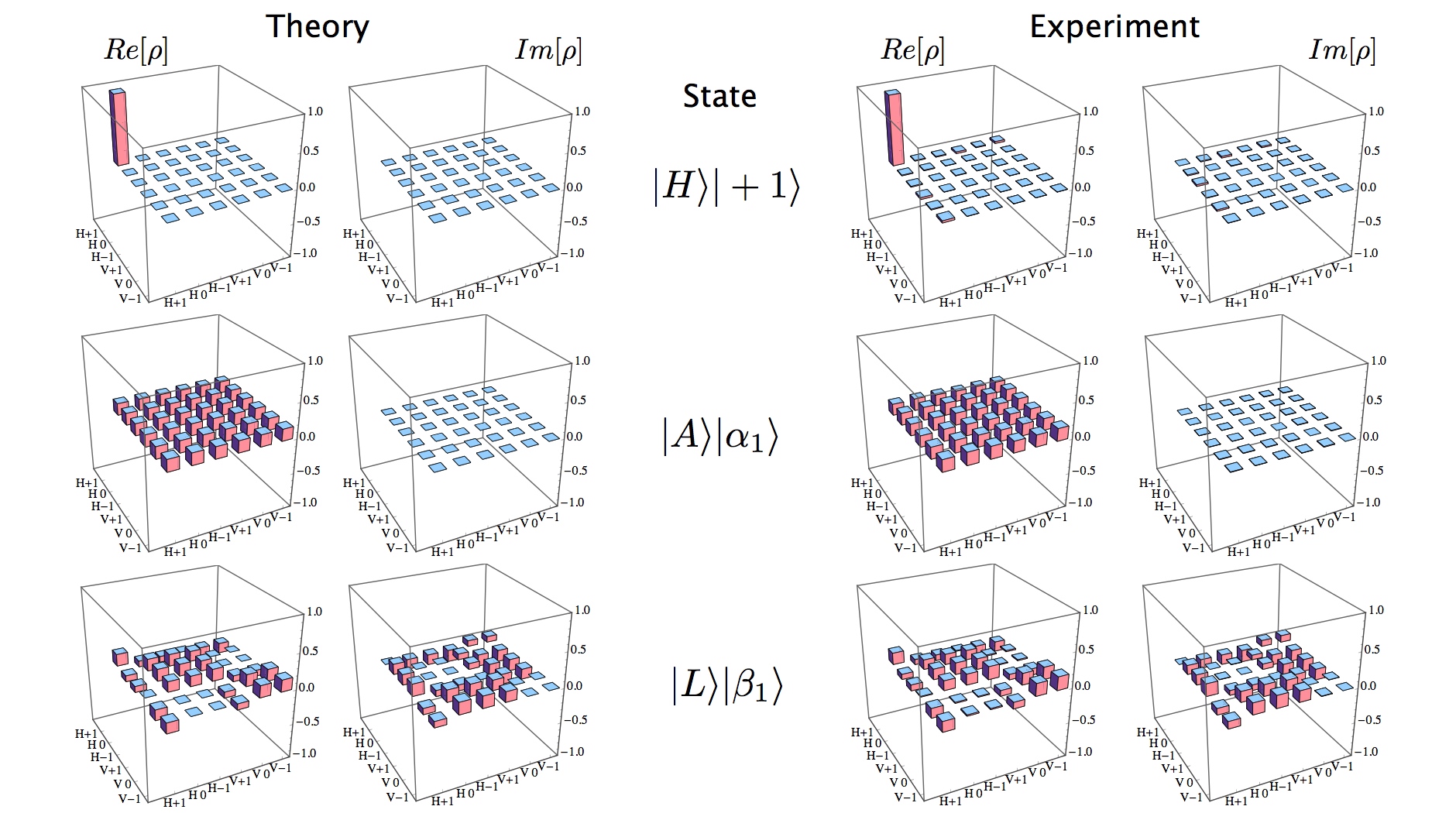}
\caption{\textbf{Quantum tomography of hybrid qusix photonic states:} Density matrices associated to states of each of the three MUBs have been fully reconstructed by projections over all the 72 states obtained by direct product of the three MUBs of the 2D polarization space $\pi$ and the four ones of the 3D OAM subspace $\mathcal{O}$. Here we show one state for each MUB. Experimental and theoretical matrices are reported for comparison.}
\label{hybrid2}
\end{figure*}

As a first test, we verified the MUBs properties by generating each qusix $\ket{\psi_i}$ among the 18 states of the MUBs and then projecting it onto all the 18 states $\ket{\psi_j}$. Figure \ref{hybrid1} shows the resulting measured probability distribution $P_{ij}=|\braket{\psi_j}{\psi_i}|^2$, compared to the theoretical one $P'_{ij}$. For a quantitative comparison, we used the similarity parameter $S=\frac{\left(\sum_{i,j}\sqrt{P_{ij}P'_{ij}}\right)^2}{\sum_{i,j}P_{ij}\sum_{i,j}P'_{ij}}$, which is a natural generalization of the fidelity used to compare two wavefunctions, finding $S=(99.19\pm0.04)\%$. As a second check of the quality of our hybrid qusix states, we reconstructed the density matrix of all the 18 states by quantum state tomography. Since we lack a complete set of MUBs in dimension six, we performed measurements in all possible product states obtained combining the three MUBs of the polarization space $\pi$ and the four MUBs in the OAM space $\mathcal{O}$, for a total of 72 projections. In Table \ref{sets}, the resulting experimental fidelities of the 18 MUBs states are reported. The overall mean fidelity was $\bar{F}=(98.51\pm 0.04)\%$. Moreover, Figure \ref{hybrid2} shows the reconstructed density matrices compared to the theoretical ones for three representative qusix states, one for each MUB considered here.

\begin{table}[h]
{\small
\begin{tabular}{|c|c|c|c|}
\hline\hline
\textbf{Basis} & \textbf{State} &  \textbf{Fidelity} \\
\hline\hline
\multirow{6}{*}{I} & $\ket{H}\ket{+1} $   & $0.986\pm0.002$\\
 			  & $\ket{H}\ket{0}$	 	    & $0.982\pm0.002$\\
			  & $\ket{H}\ket{-1}$	      & $0.986\pm0.002$\\
			  & $\ket{V}\ket{+1}$	 	    & $0.988\pm0.002$\\
			  & $\ket{V}\ket{0}$		    & $0.980\pm0.002$\\
			  & $\ket{V}\ket{-1}$	     & $0.983\pm0.002$\\
			  \hline
\multirow{6}{*}{II} & $\ket{A}\ket{\alpha_1} $  & $0.989\pm0.001$\\
 			  & $\ket{A}\ket{\alpha_2}$		    & $0.981\pm0.002$\\
			  & $\ket{A}\ket{\alpha_3}$	    & $0.986\pm0.002$\\
			  & $\ket{D}\ket{\alpha_1}$	    & $0.989\pm0.001$\\
			  & $\ket{D}\ket{\alpha_2}$	 	    & $0.982\pm0.002$\\
			  & $\ket{D}\ket{\alpha_3}$		    & $0.980\pm0.002$\\
			  \hline
\multirow{6}{*}{III} & $\ket{L}\ket{\beta_1} $  & $0.981\pm0.002$\\
 			  & $\ket{L}\ket{\beta_2}$		    & $0.981\pm0.002$\\
			  & $\ket{L}\ket{\beta_3}$	    & $0.979\pm0.002$\\
			  & $\ket{R}\ket{\beta_1}$	    & $0.977\pm0.002$\\
			  & $\ket{R}\ket{\beta_2}$	    & $0.972\pm0.002$\\
			  & $\ket{R}\ket{\beta_3}$		    & $0.970\pm0.002$\\
 \hline
&  \textbf{Average Fidelity}  & $0.9851 \pm 0.0004 $\\
 \hline\hline
\end{tabular}
}
\caption{Experimental fidelities measured for all 18 qusix hybrid states that characterize the three chosen MUBs.}
\label{sets}
\end{table}


%
%


%
%



\subsection*{PURE-OAM QUSIX ENCODING AND CHARACTERIZATION}
Our second experimental implementation of qusix photonic states has been based on the OAM space only. Although the hybrid approach may offer advantages for certain specific tasks \cite{Damb12}, an encoding in OAM is in principle suitable of extension to arbitrary dimensionality and enables the generation of any kind of state, including the entangled ones which, for hybrid encoding, would need a more complex experimental setup. To define a 6D Hilbert space, we adopted the following OAM eigenstates as logical basis:
\begin{equation}\label{logicoam}
I=\{\ket{-3},\ket{-2},\ket{-1},\ket{1},\ket{2},\ket{3}\}.
\end{equation}
The three MUBs were still defined starting from the tensor products of a 2D and a 3D spaces, as given in Eq.\ (\ref{combmubs}). More details about the resulting states of the three bases $I, II, III$ are given in Methods.

\begin{figure*}
\includegraphics[scale=.25]{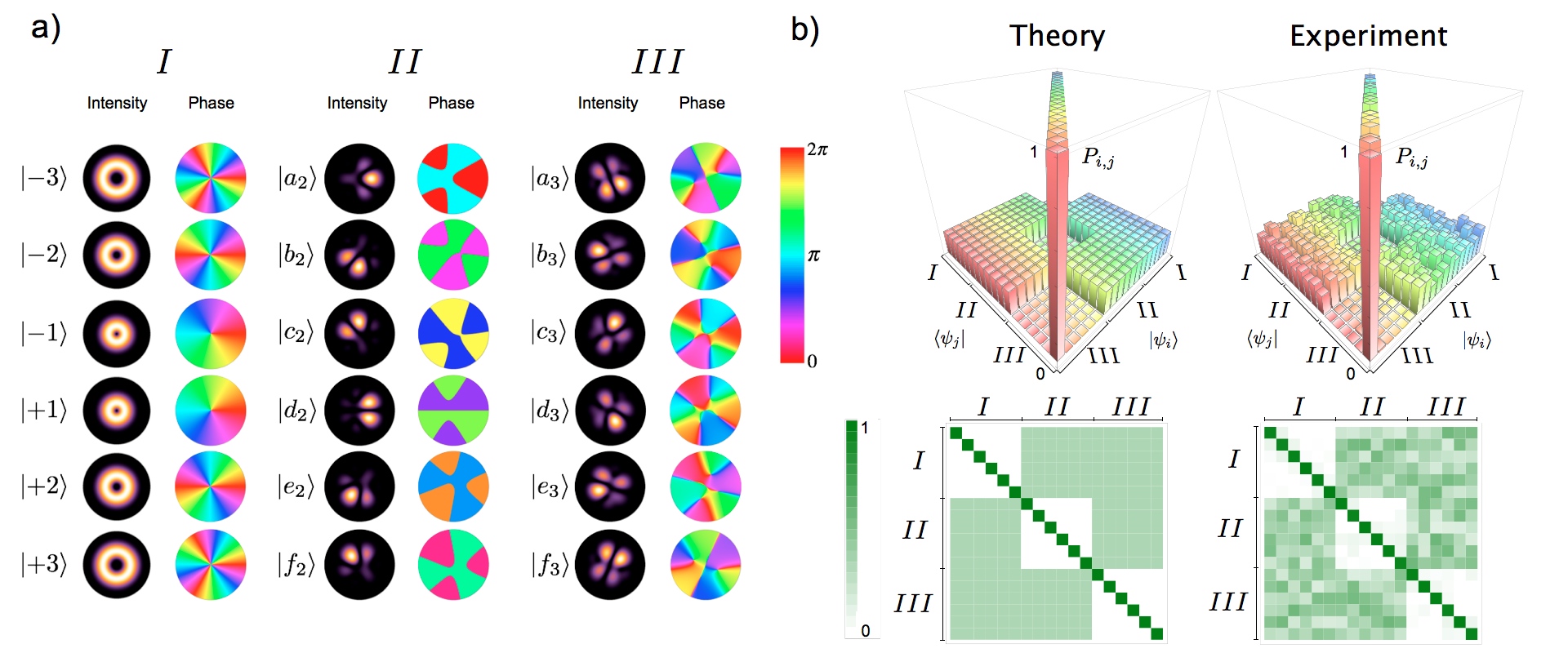}
\caption{\textbf{Experimental analysis of pure OAM qusix:} \textbf{a)} Graphical representation of all 18  states of the three selected MUBs, in the case of pure OAM 6D encoding. The precise definition of these states is given in Methods. For each state, both the intensity and phase patterns are shown. \textbf{b)} Theoretical and experimental probability distributions for an experiment in which all the $18\times18$ combinations of generated/detected states belonging to the three MUBs are tested.}
\label{pure}
\end{figure*}

The experimental setup used for generating and testing the states of the MUBs is the same as in the hybrid qusix case (see Fig.\ \ref{setup}), but with the polarization optics set so as to keep a fixed polarization everywhere. The kinoform generation was based on the method described in Sec.\ III. Figure \ref{pure}-\textbf{a} shows the intensity and phase profile of the 18 OAM modes which form the three MUBs. In Figure \ref{pure}-\textbf{b}, the theoretical and experimental probability distributions for all combinations of state preparation and detection are reported. The similarity between the two distributions is $S=(99.06\pm0.04)\%$ while the mean fidelity over the 18 states is $F=(98.78\pm0.08)\%$. Comparing this result with the hybrid case, in which only OAM states in dimension 3 were generated, we find that the fidelity of the OAM generation does not decrease rapidly with the dimensions. Hence, the holograhic method used in this work promises to be suitable for the high-fidelity generation of OAM photonic qudits with very large dimension $d$. 

\section*{DISCUSSION}
In summary, we have reported the experimental implementation of a non-extendable set of three MUBs for a photonic quantum system of dimension six by two different approaches. In the first, the qusix states have been implemented via a hybrid scheme based on polarization-orbital angular momentum encoding. All the 18 states belonging to the MUBs are in this case separable states of two different degrees of freedom. The demonstration of MUBs with high fidelity and unbiasedness has required the development of a new method for determining the kinoform to be visualized in the spatial light modulator. The second demonstrated approach was based on a quantum encoding in the photon OAM space only, at a fixed polarization. The generation of a set of MUBs with high fidelity was again verified and this method is suitable for a convenient extension to higher dimensionality. The techniques we have demonstrated here can find application in fundamental tests of quantum mechanics, quantitative wave-particle duality, quantum key distribution, and tests of quantum complementarity and logical indeterminacy.

\section*{Methods}
In dimension $d=2$, the eigenstates of the three Pauli operators provide a complete set of MUBs, which can be represented by the columns of the following three matrices:
\begin{equation}\label{pol_bases}
\pi_1\!=\!\left(
\begin{array}{cc}
1&0\\0&1
\end{array}
\right)\! , \;
\pi_2\!=\!\frac{1}{\sqrt{2}}\left(
\begin{array}{cc}
1&1\\1&-1
\end{array}
\right)\! , \;
\pi_3\!=\!\frac{1}{\sqrt{2}}\left(
\begin{array}{cc}
1&1\\ \text{i}&-\text{i}
\end{array}
\right) \!.
\end{equation}

In $d=3$, there exist four MUBs. We represent them here as the columns of the following four matrices:
\begin{eqnarray}\label{oam_bases}
\mathcal{O}_1=\left(
\begin{array}{ccc}
1&0&0\\0&1&0\\0&0&1
\end{array}
\right),\quad
\mathcal{O}_2=\frac{1}{\sqrt{3}}\left(
\begin{array}{ccc}
1&1&1\\1&\omega&\omega^2\\1&\omega^2&\omega
\end{array}
\right) \nonumber\\
\mathcal{O}_3=\frac{1}{\sqrt{3}}\left(
\begin{array}{ccc}
1&1&1\\ \omega&\omega^2&1\\\omega&1&\omega^2
\end{array}
\right),
\mathcal{O}_4=\frac{1}{\sqrt{3}}\left(
\begin{array}{ccc}
1&1&1\\ \omega^2&\omega&1\\ \omega^2&1&\omega
\end{array}
\right),
\end{eqnarray}
where $\omega=\exp{\left(\text{i} 2\pi/3 \right)}$.

In $d=6$, we may construct three MUBs by a direct product of the $\pi_1$, $\pi_2$, $\pi_3$ bases and the corresponding first three bases $\mathcal{O}_1$, $\mathcal{O}_2$, $\mathcal{O}_3$:
\begin{eqnarray}
I=\pi_1\otimes\mathcal{O}_1, \quad II=\pi_2\otimes\mathcal{O}_2, \quad III=\pi_3\otimes\mathcal{O}_3. 
\end{eqnarray}
These three 6D bases have the following matrix representation:
\begin{eqnarray}
I=\left(
\begin{array}{cccccc}
1&0&0&0&0&0\\0&1&0&0&0&0\\0&0&1&0&0&0\\0&0&0&1&0&0\\0&0&0&0&1&0\\0&0&0&0&0&1
\end{array}
\right)\\
II=\frac{1}{\sqrt{6}}\left(
\begin{array}{cccccc}\label{II}
1&1&1&1&1&1\\1&\omega&\omega^2&1&\omega&\omega^2\\1&\omega^2&\omega&1&\omega^2&\omega\\1&1&1&-1&-1&-1\\1&\omega&\omega^2&-1&-\omega&-\omega^2\\1&\omega^2&\omega&-1&-\omega^2&-\omega
\end{array}
\right)\\
III=\frac{1}{\sqrt{6}}\left(
\begin{array}{cccccc}\label{III}
1&1&1&1&1&1\\ \omega&\omega^2&1&\omega&\omega^2&1\\ \omega&1&\omega^2&\omega&1&\omega^2\\i&i&i&-i&-i&-i\\i\omega&i\omega^2&i&-i \omega&-i \omega^2&-i\\ i\omega&i&i\omega^2&-i\omega&-i&-i\omega^2
\end{array}
\right)
\end{eqnarray}
The 18 columns of these three matrices give the coefficients of the logical basis superpositions defining the 18 OAM states shown in Fig.\ \ref{pure}~\textbf{a)}.

\section*{Acknowledgments} 
This work was supported by project HYTEQ-FIRB, the Finanziamento Ateneo 2009 of Sapienza Universit\`a di Roma, the European Research Council (ERC) Starting Grant 3D-QUEST, and by the Future and Emerging Technologies (FET) program of the European Commission, under FET-Open Grant No. 255914, PHORBITECH.


\bibliographystyle{naturemag}

\end{document}